# Cross-modal Search Method of Technology Video based on Adversarial Learning and Feature Fusion

Xiangbin Liu, Junping Du*, Meiyu Liang, Ang Li

Beijing Key Laboratory of Intelligent Communication Software and Multimedia, School of Computer Science (National Pilot Software Engineering School), Beijing University of Posts and Telecommunications, Beijing 100876

**Abstract**: Technology videos contain rich multi-modal information. In cross-modal information search, the data features of different modalities cannot be compared directly, so the semantic gap between different modalities is a key problem that needs to be solved. To address the above problems, this paper proposes a novel Feature Fusion based Adversarial Cross-modal Retrieval method (FFACR) to achieve text-to-video matching, ranking and searching. The proposed method uses the framework of adversarial learning to construct a video multimodal feature fusion network and a feature mapping network as generator, a modality discrimination network as discriminator. Multi-modal features of videos are obtained by the feature fusion network. The feature mapping network projects multi-modal features into the same semantic space based on semantics and similarity. The modality discrimination network is responsible for determining the original modality of features. Generator and discriminator are trained alternately based on adversarial learning, so that the data obtained by the feature mapping network is semantically consistent with the original data and the modal features are eliminated, and finally the similarity is used to rank and obtain the search results in the semantic space. Experimental results demonstrate that the proposed method performs better in text-to-video search than other existing methods, and validate the effectiveness of the method on the self-built datasets of technology videos.

## 1 Introduction

With the rapid development of the new generation of 5G Internet, video data information is exploding [1]. More and more videos about science and technology information have an increasingly important role [2]. Technology videos mainly include academic conferences, lectures and other professional academic content, which is a valuable and rich information access channel for scientific researchers [3]. Compared with the popular short videos on the Internet, technology videos are longer, more professional, and richer in content. For scientific researchers, when they want to obtain information related to a certain field, the most common way is to use text modal search for the relevant description text of technology videos, which has high specialization requirements for the description and labeled information of technology videos, and the unlabeled videos cannot be retrieved, which is not conducive to the dissemination of technology videos and academic information. The search of single-modal data can no longer well meet the existing demand for technology video, and the demand for cross-modal information search of technology videos is increasing day by day.

In recent years, deep learning has been widely used on text and video data, and it provides support for cross-modal search by extracting data features accurately and efficiently. However, the data feature distributions between different modalities are different and the semantic spaces are not interoperable [4], and adversarial learning [5] is needed to establish the association between text and video containing the same semantic content through the semantic space. Adversarial learning is very effective for generating a new data distribution and has been widely used for text, image, and speech generation [6][7].

In this paper, we propose FFACR for searching technology videos using keyword text and ranking results according to semantic similarity. The method mainly uses adversarial learning strategy to train three neural network models, namely, video multi-modal feature fusion network, feature mapping network and modal discrimination network. The video multimodal fusion network includes pre-trained image encoder ResNet50, pre-trained text encoder Bert, and modal fusion network. The network is involved in the training process and generates a better multimodal feature representation of the video. The feature mapping network is used as a generator for adversarial learning to map features from text and video into the same semantic space, and the semantics and their similarity are used to train the feature mapping network, and the similarity constraint can reduce the difference of different modal data under the same semantics. The modality discriminator network is used as a discriminator to distinguish the original modality of the data mapped into the same semantic space. The proposed method projects the searched text into the same semantic space through the feature mapping network, and then obtains the ranking of the search results according to the semantic vector distance from the candidate data.

The proposed FFACR method is intended to solve the cross-modal search problem of technology videos, including the search of technology videos without labeled description text, and the semantic complementation and search optimization of labeled technology videos. The main contributions of this paper are as follows.

We propose a novel cross-modal search method of technology video based on adversarial learning and feature fusion to achieve text-to-video matching, ranking and searching. Through the joint learning of video multi-modal feature fusion network, feature mapping network and modal discrimination network, the text and video modal data are mapped into a unified semantic space.

We propose a feature fusion-based approach to obtain a feature representation of the multi-modal information of one video by fusing the features obtained by encoding the different modal information of the video. Compared with the single-modal information, multi-modal video features can better represent the semantic information of the video and improve the accuracy of the semantic representation of the technology video in the subspace.

We use natural semantic segmentation of video speech information to slice and dice the training samples. The speech text of the technology video has good natural segmentation information, which can guide the semantic segmentation of the video. Using automatic speech recognition algorithm, we generate speech text and then slice the long technology video into several short video samples for adversarial learning.

We use datasets of technology videos collected from the Internet, which can represent the types of videos that researchers are generally exposed to using. We use text search videos as the search task, and the results show that the FFACR algorithm can improve the cross-modal search accuracy of technology videos compared to other algorithms.

## 2 Related Work

### 2.1 Cross-modal search methods

Cross-modal search aims to solve the discrepancy problem between different modalities [23-30]. Feng et al [8] constructed a model by correlating the hidden representations of two unimodal auto encoders, trained the model by minimizing the linear combination of the errors of each modality and the correlation errors between the two modalities, and finally achieved the search, which is suitable for data with small amount of data and simple features. Hardoo et al [9] used correlation analysis of kernel methods to learn the semantic representation of data pairs. The kernel function between a sample feature and all samples needs to be computed, and the training speed is slow. Peng et al [10] proposed cross-media multiple deep network (CMDN), which learns information and cross-modal correlations within the same modality and different modalities in two successive steps. The errors brought by training will produce cumulative, inaccurate cross-modal semantic descriptions [11][12][13]. Wang et al [14] proposed a multimodal mapping method that can measure the similarity between different modal data and preserve the inter--modal and intra-modal similarity relationships. Wang et al [15] proposed a coupled linear regression framework to achieve cross-modal search and solve the regularized linear regression problem. Linear regression also under-fits for data with more complex features. Li et al [16] proposed a linear cross-modal factor analysis for simple cross-modal associations. Yao et al [17] proposed ranking canonical correlation analysis to maximize the query text and image similarity to find the common subspace. Andrew et al [18] proposed deep canonical correlation analysis as nonlinear correlation analysis. The above two methods are based on CCA (canonical correlation analysis) method, which only considers the semantic correlation between pairs of text images, and does not consider the possible correlation between different data in the same modality [19][20][21]. Zhuang et al [22] introduced dictionary learning into sparse coding for cross-modal search. This method is noise sensitive and requires high data sources. Ngiam et al [23] used multimodal learning task to learn multimodal features and used deep networks to learn cross-modal features. Complex multimodal features increase the size of the deep network, which is not suitable for large-scale data training. Yan et al [24] used deep typical correlation to analyze linear model mappings, which could not explore cross-modal data linkage well. Song et al [25] achieved fast search by XOR and bit counting operation, which saves much memory space, but increases the computational complexity of training [43][44][45].

### 2.2 Adversarial learning methods

Adversarial learning is used to obtain a generative model close to the original sample distribution by cyclically training the generator and discriminator [38-48]. Wang et al [31] used adversarial learning to solve the cross-modal search problem by cyclically training the feature mapping and modal discriminator networks. A triplet constraint is used to constrain the mapping process to minimize the distance between different modal data with the same semantics, while semantically maximizing the distance between different images and text. This method is more accurate for cross-modal semantic portrayal, but the introduction of the triplet constraint increases the amount of data training and the training speed is slow [32][33][34]. Xu et al [35] proposed deep adversarial metric learning, which introduces a deep network model for richer feature extraction, but only introduces adversarial learning to increase the cross-modal search regularization, making inter-modal semantics missing. He et al [36] proposed unsupervised cross-modal adversarial learning, which can utilize unlabeled data, but cross-modal semantics and homomodal semantics learning is more difficult, there is no uniform semantics as training targets, and it is not suitable for complex data. Li et al [37] proposed self-supervised adversarial hashing networks, using Hash coding to solve the cross-modal problem. Wu et al [38] learned Hash functions by cyclic consistency loss without paired training samples. Jiang et al [39] proposed multi-feature fusion network to solve multi-feature cross-modal problem. But the fused features come from the same modality, which can not deal with the features of the different modalities of the video [40][41][42].

## 3 The Proposed Method

The FFACR method proposed in this paper includes a video multimodal feature fusion network, a feature mapping network, and a modal discrimination network. For each text-video-semantic triad, the features of text and video are first fused and extracted, and then the text and video feature mapping networks are input separately, and the semantic vectors are input to the semantic distribution network. The semantic similarity network calculates the similarity of the input semantic vectors. The feature mapping network maps the text and video features into the common semantic space, and the modal discrimination network discriminates the modality of the data.

The feature mapping network is optimized by minimizing the semantic deviation of the same modal data and the modal deviation of the same semantic data before and after the mapping. The modal discrimination network is optimized by minimizing the error of the original modal decision of the data after mapping. The feature mapping network and the modal discriminant network are trained by adversarial learning, so that both networks eventually achieve a better structure. The overall flow is shown in Figure 1.

For each video clip, we intercept the first and last frames of the clip as the image features of this sample, and the optical character recognition (OCR) of the current frame is spliced with ASR text as the text features of this sample. All videos are pre-processed as described above to form several image-text pairs, and the videos belonging to them are used as category labels during training, together with the video description text to form the sample set.

### 3.2 Feature fusion network

For the video vector $v_i$, it needs to be obtained from the image (video frame) feature vector $i_i$ and the text feature vector $t_i$ by a multimodal feature fusion network. Denote $v_i = f_F(i_i, t_i; \theta_F)$ as feature fusion process, where $f_F(i_i, t_i; \theta_F)$ is the fusion mapping function of image and text features, and $\theta_F$ is the parameters of the fusion network. For the fusion network structure, we will try three different structures to explore the difference of feature representation performance.

Through training, feature fusion networks that can effectively fuse and represent video multimodal information can be obtained to provide good video semantic representation information for subsequent cross-modal search tasks.

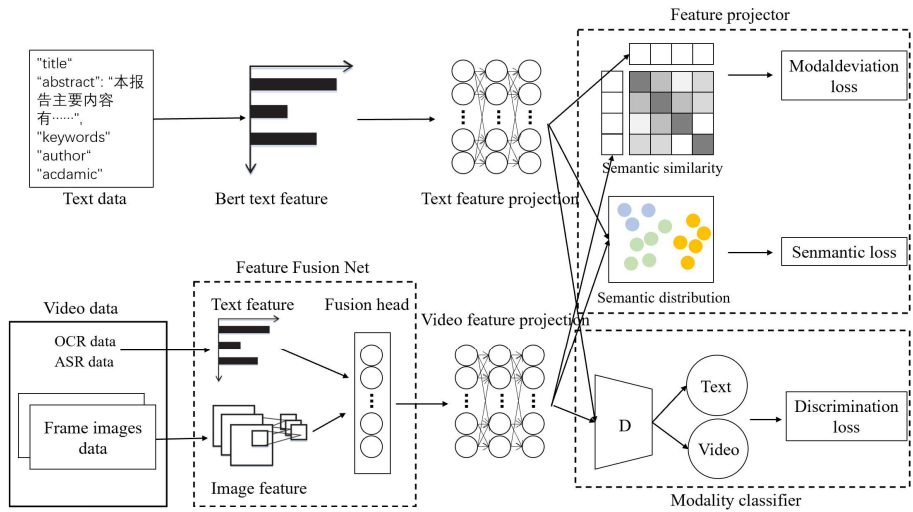

**Figure 1** FFACR network structure

### 3.1 Natural semantic video segmentation

The natural semantic video segmentation method refers to the use of automatic speech recognition (ASR) information of technology videos with timestamp information to slice a video segment (Clip) with a time period of a speaker's sentence as a training sample. Since most of the information of the technology video comes from speech, segmenting the long video according to the semantics of speech can better capture the semantic differences of the video segments.

### 3.3 Feature mapping network

In order to unify the semantics of text and video across modalities, a feature mapping network is introduced to map the data of both modalities into the common semantic space $S$ through the feature mapping network. The feature mapping of text is $S_T = f_T(T; \theta_T)$, and the feature mapping of video is $S_V = f_V(V; \theta_V)$, where $f_T(T; \theta_T)$ is the mapping function of text features, $f_V(V;\theta_V)$ is the mapping function of video features. $\theta_T$ and $\theta_V$ are the parameters representing the text feature network and the video feature network respectively. $S_T$ and $S_V$ denote the new features of text and video feature mapping in $S$ respectively, $S_V, S_T \in \mathbb{R}^{m \times n}$, the new feature dimension in the common semantic space $S$ is $m$. The FFACR method proposed in this paper obtains suitable $S_T$ and $S_V$ in the common semantic space $S$ so that they maintain the semantic relationship before mapping, and the performance of the mapping is determined by the deviation of the semantic feature distribution of the input and output contents, and the feature mapping network is trained by the difference. Under the premise of maintaining the semantic invariance, it makes the different modal data with similar semantics closer in $S$ and the different semantic data with the same modality farther in $S$.

The feature mapping network is divided into the text feature mapping network and the video feature mapping network, which are responsible for mapping the original data features into the same semantic space. In order to ensure that the mapped data maintain the original semantic features, a semantic prediction network is added after the feature mapping network, and the output of the classifier softmax is used as the result to predict the semantic distribution of the data mapped to the same semantic space. Let the parameters of the network be $\theta_{imd}$, and the $c-th$ dimensional values of the $i-th$ data semantic distribution in the text and video modalities are $p_{ic}(t_i)$ and $p_{ic}(v_i)$ respectively. And use the cross-entropy to calculate the deviation value of the semantics $L_{imd}$ in the subspace $S$, whose expression is

$$L_{imd}(\theta_{imd}) = -\frac{1}{n}\sum_{i=1}^{n}\sum_{c=1}^{d_l}(y_{ic} \times (\log p_{ic}(t_i) + \log p_{ic}(v_i))) \quad (1)$$

where $L_{imd}$ calculates the difference between the semantic distribution of each newly mapped data and the original data in the new semantic space, including the sum of the differences between the text and video components, in order to ensure that the data with similar semantics in the same modality are still close to each other in the new space $S$, and the data with farther semantics are still farther away in the new space after the transformation of $f_T(T;\theta_T)$ and $f_V(V;\theta_V)$.

In order to ensure that the data under different modalities are close to each other after feature mapping, and the data with different semantics of different modalities are far away from each other, the semantic similarity matrix $Sim_L \in \mathbb{R}^{n \times n}$ is constructed by using the semantic distribution of the original data based on the calculation of the similarity of the semantic distribution of the input content $l_{1...n}$. The semantic distributions of any two data are $l_a$ and $l_b$, respectively, and their similarity is defined as

$$sim(l_a, l_b) = \frac{l_{ai} \cdot l_{bi}}{\|l_{ai}\| \cdot \|l_{bi}\|} = \frac{\sum_{i=1}^{d_l} l_{ai} \times l_{bi}}{\sqrt{\sum_{i=1}^{d_l}(l_{ai})^2} \times \sqrt{\sum_{i=1}^{d_l}(l_{bi})^2}} \quad (2)$$

Here it is necessary to calculate the similarity for all distributions $l_{1...n}$ and obtain the semantic similarity under all data. $Sim_L$ is calculated as follows.

$$Sim_L(i,j) = sim(l_i, l_j) \quad (3)$$

In this paper, $\mathcal{L}_2$ parametric is chosen to describe the difference between two similarity matrices, and the difference value is defined as the modal deviation value $L_{imi}(\theta_T, \theta_V)$, which is calculated as shown below.

$$\begin{aligned}L_{imi}(\theta_V, \theta_T) &= \mathcal{L}_2(Sim_L, Sim_S) \\ &= \sum_{i=1}^{n}\sum_{j=1}^{n}\left\|sim(l_i, l_j) - sim(f_T(t_i;\theta_T), f_V(v_j;\theta_V))\right\|_2 \\ &= \sum_{i=1}^{n}\sum_{j=1}^{n}\left\|\frac{\sum_{k=1}^{d_l} l_{ik} \times l_{jk}}{\sqrt{\sum_{k=1}^{d_l}(l_{ik})^2} \times \sqrt{\sum_{k=1}^{d_l}(l_{jk})^2}} - \frac{\sum_{k=1}^{m} f_T(t_i;\theta_T)_k \times f_V(v_j;\theta_V)_k}{\sqrt{\sum_{k=1}^{m}(f_T(t_i;\theta_T)_k)^2} \times \sqrt{\sum_{k=1}^{d_l}(f_V(v_j;\theta_V)_k)^2}}\right\|_2\end{aligned} \quad (4)$$

The overall loss function of the feature mapping network is defined as $L_{emb}$, by a linear weighted summation of the semantic deviations $L_{imd}$ and the modal deviations $L_{imi}$, where $\alpha$ and $\beta$ denote the contribution of the two deviation values to the loss function, respectively, and the mapping loss is calculated as shown below.

$$L_{emb}(\theta_T, \theta_V, \theta_{imd}) = \alpha \cdot L_{imd} + \beta \cdot L_{imi} \quad (5)$$

where $\alpha$ and $\beta$ are used as hyperparameters of the network to determine the optimal values by subsequent experiments.

### 3.4 Modal discriminant network

The modal Discriminant network mainly distinguishes the original modality of the data mapped to the common semantic space. Let the label of the data after mapping through text be 0 and the label of the data after mapping through video be 1. The modal discrimination network determines the original modality of the data as accurately as possible. A neural network is used for the computation and the loss function of this network is defined as the deviation value of the modal prediction. The loss function $L_{adv}$ is calculated as shown below.

$$L_{adv}(\theta_D) = -\frac{1}{n}\sum_{i=1}^{n}\log(1-D(s_{ti};\theta_D)) + \log(D(s_{vi};\theta_D)) \quad (6)$$

$$= -\frac{1}{n}\sum_{i=1}^{n}\log(1-D(f_T(ti;\theta_T);\theta_D)) + \log(D(f_V(v_i;\theta_V);\theta_D))$$

where $\theta_D$ is the modal discriminant network parameter, and $D(x,\theta_D)$ represents the probability that the network determines $x$ is a video.

The training process of adversarial learning consists of the co-training of the feature mapping network and the modal discrimination network. Based on the idea of adversarial optimization of the values of the two loss functions, the goal of the feature mapping network is to maintain the semantic elimination of modality as much as possible, and the goal of the modality discrimination network is to distinguish the modality of different data in the common semantic space as much as possible. The specific training process of the feature mapping network and the modal discrimination network is shown in Algorithm 1.

---

**Algorithm 1** Training process of FFACR

**Input:** Based on mini-batch, extract the feature of the current batch data, text feature as $t_1,...,t_n$, video multi-modal feature as $v_1,...,v_n$, semantic distribution as $l_1,...,l_n$;

**Output:** Trained $\theta_T$ and $\theta_V$;

**Main iteration:** The number of iterations for feature projection network's single training is $k$, data amount is $m$ for each mini-batch, learning rate is $\mu$, loss parameter is $\lambda$;

Randomly initialize the parameters of the model;
**while** not converge **do**
  **while** $k > 0$ **do**
    Optimize $\theta_T, \theta_V$ and $\theta_{imd}$ in the direction of decreasing gradient;

$$\theta_T = \theta_T - \mu \cdot \Delta_{\theta_T} \frac{1}{m}(L_{emb} - L_{adv});$$

$$\theta_V = \theta_V - \mu \cdot \Delta_{\theta_V} \frac{1}{m}(L_{emb} - L_{adv});$$

$$\theta_{imd} = \theta_{imd} - \mu \cdot \Delta_{\theta_{imd}} \frac{1}{m}(L_{emb} - L_{adv});$$

$$k \Leftarrow k-1;$$

  **end while**
  Optimize $\theta_D$ in the direction of increasing gradient;

$$\theta_D = \theta_D + \mu \cdot \lambda \cdot \Delta_{\theta_D} \frac{1}{m}(L_{emb} - L_{adv});$$

**end while**
**return** $\theta_T$ and $\theta_V$.

---

## 4 Experimental Results and Analysis

### 4.1 Datasets

In this paper, two technology video datasets are constructed independently, and the specific information is shown in Table 1. The subject categories covered by the datasets include computer, chemistry, biology, physics, etc. The data sources include Wanfang Data Knowledge Service Platform, CCF China Software Conference video, CVPR Conference video, etc. Each sample contains the images of the first and last frame of the video clip, OCR and ASR information during the clip time period; video description information as text modality, and video number as the real category information during the training process.

**Table 1** Statistics of the datasets in our experiment

| Dataset Name | Number of Samples | Number of Videos/Labels |
|---|---|---|
| Wanfang Technology Video Dataset | 3752 | 64 |
| Technology Lecture Video Dataset | 6124 | 123 |

### 4.2 Baselines

This paper compares the proposed method with the following methods:

CCA [17]: focuses only on the association relationship between different modal data pairs, ignoring the semantic information of data categories.

JFSSL [22]: uses labels to distance the different categories of different modalities and ignores the semantic similarity of different modalities.

CMDN [10]: semantic similarity of different modalities is considered and deep neural networks are used to train the model.

ACMR [31]: uses adversarial learning to optimize the semantic distance between the same and different modalities, but uses only 0 and 1 binarization to represent the correlation between different modalities.

DSCMR [49]: text features are extracted using TextCNN, and semantic subspaces are built using deep feature mapping networks.

FFACR$_I$: FFACR method uses only video frame information.

FFACR$_T$: FFACR method uses only video text information (OCR, ASR).

### 4.3 Comparative experiments of different methods

The MAP index is used to measure the performance of different cross-modal search methods, the MAP values are calculated from the top 5, top 10 and top 30 search results for the text-to-video task on the technology video dataset. The results of the comparison methods on the technology video dataset are shown in Table 2 and 3.

As shown in Table 2 and 3, the MAP metrics of the proposed FFACR method on the technology video dataset are better than the other methods. In terms of video multi-modal feature fusion, it can be observed that the performance of using video multimodal fusion features is better than that of video single-modal features, proving the effectiveness of video multimodal fusion. And the performance using video frame single-modal features $FFACR_I$ is lower than that of text single-modal $FFACR_T$. The results are consistent with the previous assumption that the main source of technological video information is textual modality.

**Table 2** Comparison of video cross-modal search methods performance on Wanfang technology video dataset

| Dataset Name | MAP@5 | MAP@10 | MAP@30 |
|---|---|---|---|
| CCA | 0.2381 | 0.2157 | 0.2358 |
| JFSSL | 0.3510 | 0.3400 | 0.3120 |
| CMDN | 0.3608 | 0.3389 | 0.3300 |
| ACMR | 0.4603 | 0.4548 | 0.4326 |
| DSCMR | 0.4663 | 0.4322 | 0.4634 |
| FFACR | **0.4783** | **0.4820** | **0.4701** |
| $FFACR_I$ | 0.0892 | 0.0793 | 0.0854 |
| $FFACR_T$ | 0.4329 | 0.4190 | 0.4233 |

**Table 3** Comparison of video cross-modal search methods performance on technology lecture video dataset

| Dataset Name | MAP@5 | MAP@10 | MAP@30 |
|---|---|---|---|
| CCA | 0.2586 | 0.2488 | 0.2248 |
| JFSSL | 0.4923 | 0.4647 | 0.4279 |
| CMDN | 0.5444 | 0.5343 | 0.4837 |
| ACMR | 0.6875 | 0.6632 | 0.6038 |
| DSCMR | 0.6931 | 0.6729 | 0.6319 |
| FFACR | **0.7033** | **0.6734** | **0.6493** |
| $FFACR_I$ | 0.1208 | 0.1023 | 0.1190 |
| $FFACR_T$ | 0.6748 | 0.6492 | 0.6280 |

We use PR curves (precision recall curve) in Figure 2 and 3, to measure the effect of cross-modal search, which describe the relationship between precision rate and recall rate in search. Under the same recall rate, the higher the precision rate is, the less irrelevant videos are searched and the more accurate the results are.

As shown in Figures 2 and 3, the FFACR method outperforms the comparison algorithm on both datasets and is higher than $FFACR_I$ and $FFACR_T$. The gap between $FFACR_I$ and $FFACR_T$ is more obvious on the technology lecture video dataset in Figure 3. $FFACR_I$ does not effectively learn the relevant features of video data, which fits the characteristics of technology lecture video dataset: the picture is single and the text features are rich.

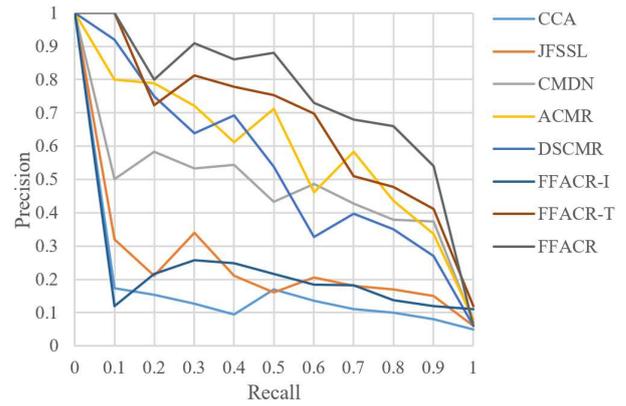

**Figure 2** PR curve on Wanfang technology video dataset

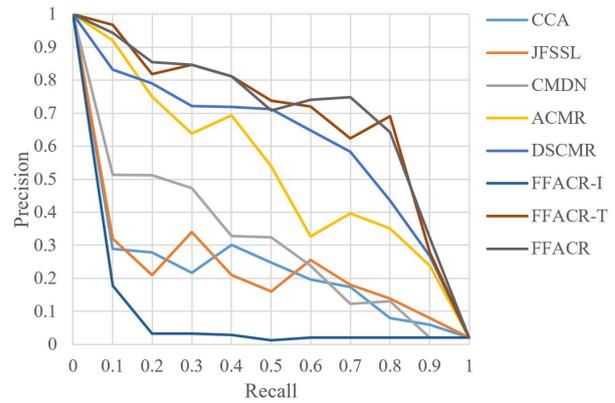

**Figure 3** PR curve on technology lecture video dataset

### 4.4 Ablation experiments

The main parameters of the FFACR method are the loss function of the feature mapping network, which are the semantic mapping loss and the weight coefficients of the modal loss $\alpha$ and $\beta$. Experiments are conducted on two technology video datasets, using the MAP values of text search videos as evaluation metrics, to determine the effect of $\alpha$ and $\beta$ values on the metrics. Two values from 0.1 to 100 in 10 times intervals get 16 sets of data, and the experimental results are shown in Figure

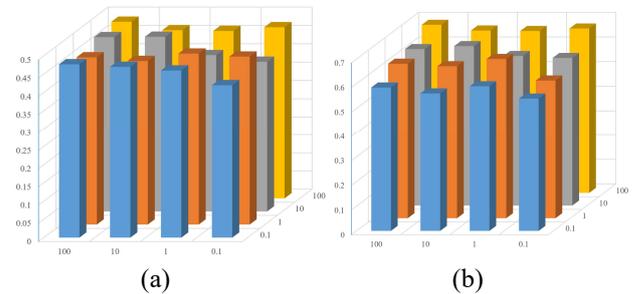

(a)      (b)

**Figure 4** (a) The performance on Wanfang technology video dataset, (b) the performance on Technology lecture video dataset, map@30. The x-axis is the parameter $\alpha$, the y-axis is the parameter $\beta$

4(a) and (b).
From Figure 4(a) and (b), we can see that FFACR performs well when $\alpha$ is in the range of [0.1, 1] and $\beta$ is in the range of [1, 10], and the MAP@30 value will not fluctuate greatly with the changes of $\alpha$ and $\beta$.

## 5 Conclusions

In this paper, we propose a novel feature fusion based adversarial cross-modal retrieval method (FFACR) to achieve text-to-video matching, ranking and searching. FFACR method using an adversarial learning-based approach to train a feature fusion network, a feature mapping network and a modal discrimination network alternately, preserving semantic distributions and eliminating modal differences. The feature fusion network gets a good extraction and fusion of video features. The feature mapping network uses semantic distributions that maintain the uniformity of the semantic distribution of the data before and after mapping. And the feature discriminant network eliminates feature differences between modalities. Experimental results on the technology video dataset show that the FFACR method has higher accuracy compared with other comparative methods, proving the effectiveness of the proposed method.

## Acknowledgements

This work was supported by the National Natural Science Foundation of China (No.62192784, No.62172056, No. 61877006).